\documentclass[times, 10pt,twocolumn]{article}
\usepackage{latex8}
\usepackage{times}

\usepackage[vlined,ruled,linesnumbered]{algorithm2e}
\usepackage[open-square,italic-theorems]{QED}
\usepackage[latin1]{inputenc}
\usepackage{epsfig}
\usepackage{subfigure}
\usepackage{latexsym}
\usepackage{amssymb}
\usepackage{pifont}

\newtheorem{thm}{Theorem}

\begin{document}

\title{Aggregating and Deploying Network Access Control Policies}

\author{
 Joaqu\'in G. Alfaro$^{~\dagger,\ddagger}$\\
 \\
 $^{\dagger}$~Universitat Oberta de Catalunya\\
Rambla Poble Nou 156, \\
08018 Barcelona - Spain\\
joaquin.garcia-alfaro@acm.org
 \and
 Fr\'ed\'eric Cuppens$^{~\ddagger}$ \hspace{1cm} Nora Cuppens-Boulahia$^{~\ddagger}$\\
 \\
 $^{\ddagger}$~GET/ENST-Bretagne,\\
 2, rue de la Ch\^ataigneraie,\\
 35576 Cesson  S\'evign\'e - France\\
 \{frederic.cuppens,nora.cuppens\}@enst-bretagne.fr
}

\maketitle
\thispagestyle{empty}
\pagestyle{empty}

\noindent {\bf Abstract.} The existence of errors or inconsistencies
in the configuration of security components, such as filtering routers
and/or firewalls, may lead to weak access control policies ---
potentially easy to be evaded by unauthorized parties. We present in
this paper a proposal to create, manage, and deploy consistent
policies in those components in an efficient way. To do so, we combine
two main approaches. The first approach is the use of an aggregation
mechanism that yields consistent configurations or signals
inconsistencies. Through this mechanism we can fold existing policies
of a given system and create a consistent and global set of access
control rules --- easy to maintain and manage by using a single
syntax. The second approach is the use of a refinement mechanism that
guarantees the proper deployment of such a global set of rules into
the system, yet free of inconsistencies.


\section{Introduction}
\label{sec:introduction}

\noindent In order to defend the resources of an information system
against unauthorized actions, a \textit{security policy} must be
defined by the administrator of an information system, i.e. a set of
rules stating what is permitted and what is prohibited in a system
during normal operations. Once specified the complete set of
prohibitions and permissions, the administrator must decide which
\textit{security mechanisms} to use in order to enforce the security
policy. This enforcement consists in distributing the security rules
expressed in this policy over different security components, such as
filtering routers and firewalls. This implies cohesion of the security
functions supplied by these components. Indeed, security rules
deployed over different components must be consistent, addressing the
same decisions under equivalent conditions, and not repeating the same
actions more than once.

A first solution to ensure these requirements is by applying a formal
security model to express network security policies. In \cite{fast},
for example, an access control language based on XML syntax and
supported by the access control model Or-BAC \cite{orbac} is proposed
for specifying access control meta-rules and, then, refined into
different firewall configuration rules through XSLT transformations.
In \cite{hassan2003}, another top-down proposal based on the RBAC
model \cite{rbac} is also suggested for such a purpose. However, and
although the use of formal models ensures cohesion, completeness and
optimization as built-in properties, in most cases, administrators are
usually reluctant to define a whole security policy from scratch, and
they expect to recycle existing configurations previously deployed
over a given system.

A second solution to guarantee consistent and non-redundant firewall
configurations consists in analyzing and fixing rules already
deployed. In \cite{al-shaer06}, for example, a taxonomy of conflicts
in security policies is presented, and two main categories are
proposed: (1) intra-firewall anomalies, which refer to those conflicts
that might exist within the local set of rules of a given firewall;
(2) inter-firewall anomalies, which refer to those conflicts that
might exist between the configuration rules of different firewalls
that match the same traffic. The authors in \cite{al-shaer06} propose,
moreover, an audit mechanism in order to discover and warn about these
anomalies. In \cite{esorics,safecomp}, we pointed out to some existing
limitations in \cite{al-shaer06}, and presented an alternative set of
anomalies and audit algorithms that detect, report, and eliminate
those intra- and inter-component inconsistencies existing on
distributed security setups --- where both firewalls and NIDSs are in
charge of the network security policy.

The main drawback of the first solution, i.e., refinement processes
such as \cite{fast, hassan2003}, relies on the necessity of formally
writing a global security policy from scratch, as well as a deep
knowledge of a given formal model. This reason might explain why this
solution is not yet widely used, and most of the times policies are
simply deployed based on the expertise and flair of security
administrators. The main drawback of the second solution, i.e., audit
processes such as \cite{al-shaer06, esorics} for analyzing local and
distributed security setups, relies on the lack of knowledge about the
deployed policy from a global point of view --- which is very helpful
for maintenance and troubleshooting tasks.

In this paper we propose to combine both solutions to better guarantee
the requirements specified for a given network access control policy.
Our procedure consists of two main steps. In the first step, the
complete set of local policies --- already deployed over each firewall
of a given system --- are aggregated, and a global security policy is
derived. It is then possible to update, analyze, and redeploy such a
global security policy into several local policies --- yet free of
anomalies --- in a further second step. We need, moreover, a previous
step for retrieving all those details of the system's topology which
might be necessary during the aggregation and deployment processes
(cf.~Section~\ref{sec:formalism}). The use of automatic network tools,
such as \cite{skybox}, may allow us to automatically generate this
information, and properly manage any change within the system.

The rest of this paper has been organized as follows. We first present
in Section~\ref{sec:formalism} the formalism we use to specify
filtering rules, an the network model we use to represent the topology
of the system. We describe in Section~\ref{sec:proposal} our
mechanisms to aggregate and deploy firewall configuration rules, and
prove the correctness of such mechanisms. We present some related work
in Section~\ref{sec:related}, and close the paper in
Section~\ref{sec:conclusions} with some conclusions and future work.

\section{Rules, Topology and Anomalies}
\label{sec:formalism}

\noindent We recall in this section some of the definitions previously
introduced in \cite{esorics,safecomp}. We first define a filtering
rule in the form $R_i: \{cnd_i\} \rightarrow decision_i$, where $i$ is
the relative position of the rule within the set of rules,
$decision_i$ is a boolean expression in $\{accept,deny\}$, and
$\{cnd_i\}$ is a conjunctive set of condition attributes
(\textit{protocol}, \textit{source}, \textit{destination}, and so on),
such that $\{cnd_i\}$ equals $A_1 \wedge A_2 \wedge ... \wedge A_p$,
and $p$ is the number of condition attributes of a given filtering
rule.

We define now a set of functions to determine which firewalls of the
system are crossed by a given packet knowing its source and
destination. Let $F$ be a set of firewalls and let $Z$ be a set of
zones. We assume that each pair of zones in $Z$ are mutually disjoint,
i.e., if $z_i \in Z$ and $z_j \in Z$ then $z_i \cap z_j = \emptyset$.
We define the predicates $connected(f_1,f_2)$ (which becomes $true$
whether there exists, at least, one interface connecting firewall
$f_1$ to firewall $f_2$) and $adjacent(f,z)$ (which becomes $true$
whether the zone $z$ is interfaced to firewall $f$). We then define a
set of paths, $P$, as follows. If $f \in F$ then $[f] \in P$ is an
atomic path. Similarly, if $[p.f_1] \in P$ (be ``$.$'' a concatenation
functor) and $f_2 \in F$, such that $f_2 \notin p$ and
$connected(f_1,f_2)$, then $[p.f_1.f_2] \in P$. Let us now define
functions $first$, $last$, and $tail$ from $P$ in $F$ such that if $p$
is a path, then $first(p)$ corresponds to the first firewall in the
path, $last(p)$ corresponds to the last firewall in the path, and
$tail(f,p)$ corresponds to rest of firewalls in the path after
firewall $f$. We also define the order functor between paths as $p_1
\leq p_2$, such that path $p_1$ is shorter than $p_2$, and where all
the firewalls within $p_1$ are also within $p_2$. We define functions
$route$ such that $p \in route(z_1,z_2)$ iff path $p$ connects zone
$z_1$ to zone $z_2$, i.e., $p \in route(z_1,z_2)$ iff
$adjacent(first(p),z_1)$ and $adjacent(last(p),z_2)$; and
$minimal\_route$ (or $MR$ for short), such that $p \in MR(z_1,z_2)$
iff the following conditions hold: (1) $p \in route(z_1,z_2)$; (2)
there does not exist $p' \in route(z_1,z_2)$ such that $p' < p$.

Let us finally close this section by overviewing the complete set
of anomalies defined in our previous work \cite{esorics,safecomp}:\\

\noindent \textbf{Intra-firewall anomalies}
  \begin{itemize}
  \item \textbf{Shadowing -- } A configuration rule $R_i$ is shadowed
    in a set of configuration rules $R$ whether such a rule never
    applies because all the packets that $R_i$ may match, are
    previously matched by another rule, or combination of rules, with
    higher priority.

  \item \textbf{Redundancy --} A configuration rule $R_i$ is redundant
    in a set of configuration rules $R$ whether the following
    conditions hold: (1) $R_i$ is not shadowed by any other rule or
    set of rules; (2) when removing $R_i$ from $R$, the security
    policy does not change.
  \end{itemize}

\noindent \textbf{Inter-firewall anomalies}
  \begin{itemize}
  \item \textbf{Irrelevance --} A configuration rule $R_i$ is
    irrelevant in a set of configuration rules~$R$ if one of the
    following conditions holds: (1) Both source and destination
    address are within the same zone; (2) The firewall is not within
    the minimal route that connects the source zone to the destination
    zone.

  \item \textbf{Full/Partial-redundancy --} A redundancy
    anomaly\footnote{Although this kind of redundancy is sometimes
      expressly introduced by network administrators (e.g., to
      guarantee the forbidden traffic will not reach the destination),
      it is important to report it since, if such a rule is applied,
      we may conclude that at least one of the redundant components is
      wrongly working.} occurs between two firewalls whether the
    firewall closest to the destination zone blocks (completely or
    partially) traffic that is already blocked by the first firewall.

  \item \textbf{Full/Partial-shadowing --} A shadowing anomaly occurs
    between two firewalls whether the one closest to the destination
    zone does not block traffic that is already blocked by the first
    firewall.

  \item \textbf{Full/Partial-misconnection --} A misconnection a\-noma\-ly
    occurs between two firewalls whether the closest firewall to the
    source zone allows all the traffic --- or just a part of it --- that
    is denied by the second one.
  \end{itemize}

\section{Proposed Mechanisms}
\label{sec:proposal}

\noindent The objective of our proposal is the following. From a set
$F$ of firewalls initially deployed over a set $Z$ of zones, and if
neither intra- nor inter-firewall anomalies apply over such a setup,
we aim to derive a single global security police setup $R$, also free
of anomalies. Then, this set of rules $R$ can be maintained and
updated\footnote{These operations are not
  covered in the paper.} as a whole, as well as redeployed over the
system through a further refinement process. We present in the
following the main processes of our proposal.

\subsection{Aggregation of Policies}
\label{sec:aggregation}

\noindent Our aggregation mechanism works as follows. During an
initial step (not covered in this paper) it gathers all those details
of the system's topology which might be necessary during the rest of
stages. The use of network tools, such as \cite{skybox}, allows us to
properly manage this information, like the set $F$ of firewalls, the
set of configurations rules $f[rules]$ of each firewall $f \in F$, the
set $Z$ of zones of the system, and some other topological details
defined in Section \ref{sec:formalism}. An analysis of intra-firewall
anomalies is then performed within the first stage of the aggregation
process, in order to discover and fix any possible anomaly within the
local configuration of each firewall $f \in F$. In the next step, an
analysis of inter-firewall anomalies is performed at the same time
that the aggregation of polices into $R$ also does. If an anomaly
within the initial setup is discovered, then an aggregation error
warns the officer and the process quits. Conversely, if no
inter-firewall anomalies are found, then a global set of rules $R$ is
generated and so returned as a result of the whole aggregation
process.

\begin{algorithm}
  \caption{\texttt{aggregation}($F$)}
  \label{alg:merging}
  \SetKwFunction{Source}{source}
  \SetKwFunction{Destination}{destination}
  \SetKwFunction{First}{first}
  \SetKwFunction{Next}{tail}
  \SetKwFunction{Empty}{empty}
  \SetKwFunction{isFirewall}{isFirewall}
  \SetKwFunction{MinimalRoute}{MR}
  \SetKwFunction{Error}{aggregationError}
  \SetKwFunction{Exit}{exit}
  \SetKwFunction{Empty}{empty}
  \SetKwFunction{Exclusion}{exclusion}
  \SetKwFunction{PolicyRewriting}{policy-rewriting}
  \texttt{/*Phase 1*/}\\
  \ForEach{$f_1 \in F$}{\label{st:first-beg}
    \PolicyRewriting($f_1[rules]$);
  }
  \texttt{/*Phase 2*/}\\
  $R~\leftarrow~\emptyset$;\\
  $i~\leftarrow~\emptyset$;\\
  \ForEach{$f_1 \in F$}{\label{st:first-end}
    \ForEach{$r_1 \in f_1[rules]$}{
      {\footnotesize
      $Z_s \leftarrow~ \{ z \in Z~|~z~\cap$ \Source($r_1$) $\neq~\emptyset\}$;\\
      $Z_d \leftarrow~\{ z \in Z~|~z~\cap$ \Destination($r_1$)
      $\neq~\emptyset\}$;\\
    }
      \ForEach{$z_1 \in Z_s$}{
        \ForEach{$z_2 \in Z_d$}{
          \uIf{($z_1 = z_2$) \textbf{or} ($f_1 \notin
            \MinimalRoute(z_1,z_2)$)\label{if:irrelevance}}{
            \Error();\\
            \textbf{return} $~~\emptyset$;\label{if:irrelevance-return}
          }
          \uElseIf{($r_1[decision]=$``$accept$'')}{
            \ForEach{{\scriptsize$f_2 \in \MinimalRoute(z_1,z_2)$}}{
                {\scriptsize
                  $f_{2}rd \leftarrow \emptyset$;\\
                  $f_{2}rd \leftarrow \{ r_2 \in f_2~|~r_1 \backsim
                  r_2~\land $\\
                  $~~~~~r_2[decision] = $``$deny$''$\}$;\\
                }
                \eIf{($\lnot\Empty(f_{2}rd)$)\label{if:misconnection}}{
                  \Error();\\
                  \textbf{return} $~~\emptyset$;
                }{
                  {\scriptsize
                  $f_{2}ra \leftarrow \emptyset$;\\
                  }
                  {\scriptsize
                    $f_{2}ra \leftarrow \{ r_2 \in f_2~|~r_1 \backsim
                    r_2~\land $\\
                  }
                  {\scriptsize
                    $~~~~~r_2[decision] = $``$accept$''$\}$;\\
                  }
                  \ForEach{$r_2 \in f_{2}ra$}{
                    {\footnotesize
                    $R_i \leftarrow R_i \cup r_2$;\label{union1}\\
                    $R_i[source] \leftarrow z_1$;\\
                    $R_i[destination] \leftarrow z_2$;\\
                    $i \leftarrow (i+1)$;\\
                    $r_2 \leftarrow \emptyset$;\\
                    }
                  }
                }
              }
          }
          \uElseIf{{\scriptsize($f_1=$\First(\MinimalRoute($z_1,z_2$)))\label{if:refundancy}}}{
              $f_{3}r \leftarrow \emptyset$;\\
              \ForEach{{\scriptsize$f_3 \in \Next(f_1,\MinimalRoute(z_1,z_2))$}}{
                {\scriptsize
                  $f_{3}r \leftarrow \{ r_3 \in f_3 | r_1 \backsim
                  r_3\} \cup f_{3}r$;
                }
              }
              \eIf{($\lnot \Empty(f_{3}r)$\label{if:shadowing})}{
                \Error();\\
                \textbf{return} $~~\emptyset$;
              }{
                $R_i \leftarrow R_i~\cup~r_1$;\label{union2}\\
                $R_i[source] \leftarrow z_1$;\\
                $R_i[destination] \leftarrow z_2$;\\
                $i \leftarrow (i+1)$;\\
                $r_1 \leftarrow \emptyset$;\\
              }
          }
          \Else{
            \Error();\\
            \textbf{return} $~~\emptyset$;
          }
        }
      }
    }
  }
  \PolicyRewriting($R$);\\
  \Return{ $R$;}
\end{algorithm}

We present in Algorithm~1 our proposed aggregation process. The input
data is a set $F$ of firewalls whose configurations we want to fold
into a global set of rules $R$. For reasons of clarity, we assume in
our algorithm that one can access the elements of a set as a
linked-list through the operator $element_i$. We also assume one can
add new values to the list as any other normal variable does
($element_i \leftarrow value$), as well as to both remove and
initialize elements through the addition of an empty set ($element_i
\leftarrow \emptyset$). The internal order of elements from the
linked-list, moreover, keeps with the relative ordering of elements.

The aggregation process consists of two main phases. During the first
phase (cf. lines~2 and~3 of Algorithm~1), and through an iterative
call to the auxiliary function $policy$-$rewriting$ (cf.~Algorithm~4),
it analyzes the complete set $F$ of firewalls, in order to discover
and remove any possible intra-firewall anomaly. Thus, after this first
stage, no useless rules in the local configuration of any firewall $f
\in F$ might exist. We refer to Section~\ref{sec:rewriting} for a more
detailed description of this function.

During the second phase (cf. lines 5--51 of Algorithm~1), the
aggregation of firewall configurations is performed as follows. For
each permission configured in a firewall $f \in F$, the process folds
the whole chain\footnote{The operator ``$\backsim$'' is used within
  Algorithm~1 to denote that two rules $r_i$ and $r_j$ are correlated
  if every attribute in $r_i$ has a non empty intersection with the
  corresponding attribute in $r_j$.} of permissions within the
components on the minimal route from the source zone to the
destination zone; and for each prohibition, it directly keeps such a
rule, assuming it becomes to the closest firewall to the source, and
no more prohibitions should be placed on the minimal route from the
source zone to the destination zone. Moreover, and while the
aggregation of policies is being performed, an analysis of
inter-firewall anomalies is also applied in parallel. Then, if any
inter-firewall anomaly is detected during the aggregation of rules $R
\leftarrow aggregation(F)$, a message of error is raised and the
process quits.

Let us for example assume that during the aggregation process, a
filtering rule $r_i \in f_i[rules]$ presents an inter-firewall
irrelevance, i.e., $r_i$ is a rule that applies to a source zone $z_1$
and a destination zone $z_2$ (such that $s = z_1 \cap source(r_i) \neq
\emptyset$, $d = z_2 \cap destination(r_i) \neq \emptyset$) and either
$z_1$ and $z_2$ are the same zone, or firewall $f_i$ is not in the
path $[f_1,f_2,...,f_k] \in MR(z_1,z_2)$. In this case, we can observe
that during the folding process specified by Algorithm~1, the
statement of line~13, i.e., $(z_1 = z_2)~or~(f_i \notin MR(z_1,z_2))$,
becomes $true$ and, then, the aggregation process finishes with an
error and returns an empty set of rules (cf. statements of lines~14
and 15). Similarly, let us assume that $r_i \in f_i[rules]$ presents
an inter-firewall redundancy, i.e., $r_i$ is a prohibition that
applies to a source zone $z_1$ and a destination zone $z_2$ (such that
$s = z_1 \cap source(r_i) \neq \emptyset$, $d = z_2 \cap
destination(r_i) \neq \emptyset$, and $[f_1,f_2,...,f_k] \in
MR(z_1,z_2)$) and firewall $f_i$ is not the first component in
$MR(z_1,z_2)$. In this case, we can observe that during the folding
process specified by Algorithm~1, the statement of line~34, i.e., $f_i
= first(MR(z_1,z_2))$, becomes $false$ and, then, the aggregating
process finishes with an error and returns an empty set of rules.

Let us now assume that $r_i \in f_i[rules]$ presents an inter-firewall
shadowing, i.e., $r_i$ is a permission that applies to a source zone
$z_1$ and a destination zone $z_2$ such that there exists an
equivalent prohibition $r_j$ that belongs to a firewall $f_j$ which,
in turn, is closer to the source zone $z_1$ in $MR(z_1,z_2)$. In this
case, we can observe that during the folding process specified by
Algorithm~1, the statement of line~38 detects that, after a
prohibition in the first firewall of $MR(z_1,z_2)$, i.e., $f_j =
first(MR(z_1,z_2))$, there is, at least, a permission $r_i$ that
correlates the same attributes. Then, the aggregating process finishes
with an error and returns an empty set of rules. Let us finally assume
that $r_i \in f_i[rules]$ presents an inter-firewall misconnection,
i.e., $r_i$ is a prohibition that applies to a source zone $z_1$ and a
destination zone $z_2$ such that there exists, at least, a permission
$r_j$ that belongs to a firewall $f_j$ closer to the source zone $z_1$
in $MR(z_1,z_2)$. In this case, we can observe that during the folding
process specified by Algorithm~1, the statement of line~21 detects
this anomaly and, then, the process finishes with an error and returns
an empty set of rules.

It is straightforward then to conclude that whether no inter-firewall
anomalies apply to any firewall $f \in F$, our aggregation process
returns a global set of filtering rules $R$ with the union of all the
filtering rules previously deployed over $F$. It is yet necessary to
perform a post-process of $R$, in order to avoid the redundancy of all
permissions, i.e., $accept$ rules, gathered during the aggregating
process. In order to do so, the aggregation process calls at the end
of the second phase (cf.~line 50 of Algorithm~1) to the auxiliary
function $policy$-$rewriting$ (cf.~Algorithm~4). We offer in the
following a more detailed description of this function.


\subsection{Policy Rewriting}
\label{sec:rewriting}

\noindent We recall in this section our audit process to discover and
remove rules that never apply or are redundant in local firewall
policies \cite{phoenix,ssi}. The process is based on the analysis of
relationships between the set of configuration rules of a local
policy. Through a rewriting of rules, it derives from an initial set
$R$ to an equivalent one $Tr(R)$ completely free of dependencies
between attributes, i.e., without either redundant or shadowed rules.
The whole process is split in three main functions (cf. algorithms~2,
3 and 4).

The first function, $exclusion$ (cf.~Algorithm~2), is an auxiliary
process which performs the exclusion of attributes between two rules.
It receives as input two rules, $A$ and $B$, and returns a third rule,
$C$, whose set of condition attributes is the exclusion of the set of
conditions from $A$ over $B$. We represent the attributes of each rule
in the form of $Rule[cnd]$\footnote{We use
  the notation $A_i$ and $B_i$ as an abbreviation of both $A[cnd][i]$
  and $B[cnd][i]$ during the statements of lines 6--12.} as a boolean
expression over $p$ possible attributes (such as source, destination,
protocol, ports, and so on). Similarly, we represent the decision of
the rule in the form $Rule[decision]$ as a boolean variable whose
values are in $\{accept,deny\}$. Moreover, we use two extra elements
for each rule, in the form $Rule[shadowing]$ and $Rule[redundancy]$,
as two boolean variables in $\{true,false\}$ to store the reason for
why a rule may disappear during the process.

The second function, $testRedundancy$ (cf.~Algorithm~3), is a boolean
function in $\{true,false\}$ which, in turn, applies the
transformation $exclusion$ (cf. Algorithm~2) over a set of
configuration rules to check whether the first rule is redundant,
i.e., applies the same policy, regarding the rest of rules.

Finally, the third function, $policy$-$rewriting$ (cf.~Algorithm~4),
performs the whole process of detecting and removing the complete set
of intra-firewall anomalies. It receives as input a set $R$ of rules,
and performs the audit process in two different phases.

\begin{algorithm}
\caption{\texttt{exclusion}($B$,$A$)}
\label{alg:exclusion}
    $C[cnd] \leftarrow \emptyset$;\\
    $C[decision] \leftarrow B[decision]$;\\
    $C[shadowing] \leftarrow false$;\\
    $C[redundancy] \leftarrow false$;\\
    \ForAll{\small{\textbf{the elements of} $A[cnd]$ \textbf{and} $B[cnd]$}}{
        \eIf{$((A_1 \cap B_1) \neq \emptyset$ {\bf and}
          $(A_2 \cap B_2) \neq \emptyset$ {\bf and} ... {\bf and}
          $(A_p \cap B_p) \neq \emptyset)$}{
          $C[cnd] \leftarrow C[cnd]~\cup$ \\
          \footnotesize{
          \{$(B_1 - A_1) \wedge B_2 \wedge ... \wedge B_p$,\\
          $(A_1 \cap B_1) \wedge (B_2 - A_2) \wedge ... \wedge B_p$,\\
          $(A_1 \cap B_1) \wedge (A_2 \cap B_2) \wedge (B_3 - A_3) \wedge ... \wedge B_p$,\\
          $ ... $\\
          $(A_1 \cap B_1) \wedge ... \wedge (A_{p-1} \cap B_{p-1}) \wedge (B_p -
          A_p) \}$;\\
          }
        }{
          $C[cnd] \leftarrow (C[cnd] \cup B[cnd]$);
        }
    }
    \Return{
      $C$;
}
\end{algorithm}
\begin{algorithm}
  \caption{\texttt{testRedundancy}($R$,$r$)}
  \label{alg:testRedundancy}
  \SetKwFunction{Exclusion}{exclusion}
  \SetKwFunction{Warning}{warning}
    $i \leftarrow 1$;\\
    $temp \leftarrow r$;\\
    \While{$\neg test$ {\bf and} ($i \leq count(R)$)}{
        $temp \leftarrow \Exclusion(temp,R_i)$;\\
        \If{$temp[cnd]$ = $\emptyset$}{
          \Return{ $true$;}\\
        }
      $i \leftarrow (i+1)$;\\
    }
    \Return{ $false$; }
\end{algorithm}
\begin{algorithm}[h]
  \caption{\texttt{policy-rewriting}($R$)}
  \label{alg:intra-firewall-audit}
  \SetKwFunction{Exclusion}{exclusion}
  \SetKwFunction{TestRedundancy}{testRedundancy}
    $n \leftarrow count(R)$;\\
    \texttt{/*Phase 1*/}\\
    \For{$i \leftarrow 1$ \KwTo $(n-1)$}{
      \For{$j \leftarrow (i+1)$ \KwTo $n$}{
        \If{$R_i[decision] \neq R_j[decision]$}{
          ~~$R_j \leftarrow$ \Exclusion($R_j$,$R_i$);\\
          ~~\lIf{$R_j[cnd] = \emptyset$}{\\
            ~~$R_j[shadowing] \leftarrow true$; }
        }
      }
    }
    \texttt{/*Phase 2*/}\\
    \For{$i \leftarrow 1$ \KwTo $(n-1)$}{
      $R_a \leftarrow \{ r_k \in R~|~n \ge k>i$ {\bf and}\\~~~~~$r_k[decision] = r_i[decision]\}$;\\
      \eIf{\TestRedundancy($R_a$,$R_i$)}{ $R_i[cnd] \leftarrow
        \emptyset$;\\ $R_i[redundancy] \leftarrow true$; }{ \For{$j
          \leftarrow (i+1)$ \KwTo $n$}{
          \If{$R_i$[decision]=$R_j$[decision]}{
            $R_j\leftarrow$\Exclusion($R_j$,$R_i$);\\
            \lIf{($\neg R_j[redundancy]$ \textbf{and}\\
              $R_j[cnd]$ = $\emptyset$)}{\\
              ~~$R_j[shadowing] \leftarrow true$; } } } } }
\end{algorithm}

During the first phase, any possible shadowing between rules with
different decision values is marked and removed by iteratively
applying function $exclusion$ (cf. Algorithm~2). The resulting set of
rules obtained after the execution of the first phase is again
analyzed when applying the second phase.

Each rule is first analyzed, through a call to function
$testRedundancy$ (cf. Algorithm~3), to those rules written after the
checked rule but that can apply the same decision to the same traffic.
If such a test of redundancy becomes $true$, the rule is marked as
redundant and then removed. Otherwise, its attributes are then
excluded from the rest of equivalent rules but with less priority in
the order. In this way, if any shadowing between rules with the same
decision remained undetected during the first phase, it is then marked
and removed.

Based on the processes defined in algorithms~2,~3, and~4, we can
prove\footnote{A set of proofs to validate Theorem~1, as well as a
  complexity analysis of function $policy$-$rewriting$
  (cf.~Algorithm~4) and its performance in a research prototype, is
  provided in \cite{phoenix}.} the following theorem:

\begin{thm}\label{thm:intra-fw-correctness-theorem}
  Let $R$ be a set of filtering rules and let $Tr(R)$ be the resulting
  filtering rules obtained by applying Algorithm 4 to $R$. Then the
  following statements hold: (1) $R$ and $Tr(R)$ are equivalent; (2)
  Ordering the rules in $Tr(R)$ is no longer relevant; (3) $Tr(R)$ is
  free from both shadowing and redundancy.\\
\end{thm}

\subsection{Deployment of Rules}
\label{sec:deployment}

\noindent We finally present in Algorithm~5 our proposed refinement
mechanism for the deployment of an updated global set of rules. The
deployment strategy defined in the algorithm is the following. Let $F$
be the set of firewalls that partitions the system into the set $Z$ of
zones. Let $R$ be the set of configuration rules resulting from the
maintenance of a given global set of rules obtained from the
aggregation process presented in Section~\ref{sec:aggregation}
(cf.~Algorithm~1). Let $r \in R$ be a configuration rule that applies
to a source zone $z_1$ and a destination zone $z_2$, such that $s =
z_1 \cap source(r) \neq \emptyset$ and $d = z_2 \cap destination(r)
\neq \emptyset$. Let $r'$ be a rule identical to $r$ except that
$source(r') =s$ and $destination(r') =d$. Let us finally assume that
$[f_1, f_2, \ldots, f_k] \in MR(z_1,z_2)$. Then, any rule $r \in R$ is
deployed over the system as follows:

\begin{itemize}
\item If $r[decision]= accept$ then deploy a permission $r'$ on every
  firewall on the minimal route from source $s$ to destination $d$.

\item If $r[decision]= deny$ then deploy a single\footnote{This
    decision is a choice for avoiding inter-firewall redundancy in the
    resulting setup.} prohibition $r'$ on the most-upstream firewall
  (i.e., the closest firewall to the source) of the minimal route from
  source $s$ to destination $d$. If such a firewall does not exist,
  then generate a deployment error message.
\end{itemize}

\begin{algorithm}
  \caption{\texttt{deployment}($R$,$Z$)}
  \label{alg:deployment}
  \SetKwFunction{Source}{source}
  \SetKwFunction{Destination}{destination}
  \SetKwFunction{First}{first}
  \SetKwFunction{MinimalRoute}{MR}
  \SetKwFunction{Error}{deploymentError}
  \SetKwFunction{Exit}{exit}
  \SetKwFunction{Empty}{empty}
  \SetKwFunction{PolicyRewriting}{policy-rewriting}
    \PolicyRewriting($R$);\\
    \ForEach{$r_1 \in R$}{
      $Z_s \leftarrow~ \{ z \in Z~|~z~\cap$ \Source($r_1$) $\neq~\emptyset\}$;\\
      $Z_d \leftarrow~\{ z \in Z~|~z~\cap$ \Destination($r_1$) $\neq~\emptyset\}$;\\
        \ForEach{$z_1 \in Z_s$}{
          \ForEach{$z_2 \in Z_d$}{
            \uIf{$r_1[decision]=$``$accept$''}{
              \ForEach{$f_1 \in \MinimalRoute(z_1,z_2)$}{
                  $r_1' \leftarrow r$;\\
                  $r_1'[source] \leftarrow Z_1$;\\
                  $r_1'[destination] \leftarrow Z_2$;\\
                  $f_1[rules] \leftarrow f_1[rules] \cup r'$;
              }
            }
            \ElseIf{$r_1[decision]=$``$deny$''}{
              $f_1 \leftarrow~ \First(\MinimalRoute(z_1,z_2))$;\\
                \eIf{($\lnot$\Empty($f_1$))}{
                  $r_1' \leftarrow r$;\\
                  $r_1'[source] \leftarrow Z_1$;\\
                  $r_1'[destination] \leftarrow Z_2$;\\
                  $f_1[rules] \leftarrow f_1[rules] \cup r'$;\\
                }{
                  \Error();\\
                  \Exit();
                }
              }
          }
        }
      }
\end{algorithm}

It is straightforward now to prove that the deployment of a given set
of rules $R$ through Algorithm~5 is free of either intra- and/or
inter-firewall anomalies (cf.~Section~\ref{sec:formalism}). On the one
hand, during the earliest stage of Algorithm~5, the complete set of
rules in $R$ is analyzed and, if necessary, fixed with our
$policy$-$rewriting$ process (cf.~Section~\ref{sec:rewriting},
Algorithm~4). Then, by Theorem~\ref{thm:intra-fw-correctness-theorem},
we can guarantee that neither shadowed nor redundant rules might exist
in $R$. Moreover, it also allows us to guarantee that the order
between rules in $R$ is not relevant. On the other hand, the use of
the deployment strategy defined above allows us to guarantee that the
resulting setup is free of inter-firewall anomalies. First, since each
permission $r_a$ in $R$ opens a flow of permissions over all the
firewalls within the minimal routes from the source to the destination
pointed by $r_a$, and since any other rule $r'$ in $R$ cannot match
the same traffic that $r_a$ matches, we can guarantee that neither
inter-firewall shadowing nor inter-firewall misconnection can appear
in the resulting setup. Second, since each prohibition $r_d$ in $R$ is
deployed just once in the closest firewall to the source pointed by
$r_d$, and since any other rule $r'$ in $R$ cannot match the same
traffic that $r_d$ matches, we can guarantee that any inter-firewall
redundancy can appear in the resulting setup.

\section{Related Work}
\label{sec:related}

\noindent A first solution to deploy access control policies free of
errors is by applying a refinement mechanism. Hence, following such a
top-down mechanism, one can deploy a global security policy into
several component's configurations \cite{fast, bartal, hassan2003}.

In \cite{fast}, for example, a formal approach based on the Or-BAC
model \cite{orbac} is presented for this purpose. There, a set of
filtering rules, whose syntax is specific to a given firewall, may be
generated using a transformation process. The authors in
\cite{bartal}, on the other hand, use the concept of roles to define
network capabilities and refinement of policies. Indeed, they propose
the use of an inheritance mechanism through a hierarchy of entities to
automatically generate permissions.

However, their work does not fix, from our point of view, clear
semantics, and their concept of role becomes ambiguous as we pointed
out in \cite{fast}. Another work based on policy refinement is the
RBNS model \cite{hassan2003}. However, and although the authors claim
that their work is based on the RBAC model \cite{rbac}, it seems that
they only keep from this model only the concept of role. Indeed, the
specification of network entities and role and permission assignments
are not rigorous and does not fit any reality \cite{fast}.

The use of these refinement proposals \cite{fast, bartal, hassan2003}
ensures cohesion, completeness and optimization as built-in
properties. However, it is not always enough to ensure that the
firewall configuration is completely free of errors and, often,
administrators are reluctant to follow such a proposal. For this
reason, we extended in this paper the approach presented in
\cite{fast}, offering to administrators the possibility of aggregating
existing configurations before moving to such a refinement approach.


Support tools, on the other hand, are intended to directly assist
administrators in their task of configuring from scratch firewall
configurations. Firewall Builder \cite{kurland03}, for example,
provides a user interface to be used to specify a network access
control policy and then this policy is automatically translated into
various firewall configuration languages such as NetFilter
\cite{netfilter}, IpFilter \cite{ipfilter} or Cisco PIX
\cite{ciscopix}. It also provides higher portability. For instance, if
in a given network infrastructure, IpFilter is replaced by NetFilter,
it will not be necessary to completely reconfigure NetFilter. Firewall
Builder will automatically generate the rules necessary to configure
this firewall.

However, we observed some problems when using Fiwerall Builder. First,
we noticed that it might generate incorrect rules. In the case of
NetFilter, for example, we experienced the generation of rules
associated to FORWARD when they should be associated to OUTPUT and
INPUT chains. Second, we noticed the generation of redundant rules,
although such redundancy was not specified within the policy. Third,
it includes a mechanism called {\em shadowing} to detect redundancy in
the policy. However, this shadowing mechanism only detects simple
redundancy that corresponds to trivial equality or inclusion between
zones. More complex redundancies (as the anomalies defined in
~Section~\ref{sec:formalism}) are unfortunately not
detected. 

Some other proposals, such as \cite{al-shaer06, fireman, esorics,
  safecomp}, provide means to directly manage the discovery of
anomalies from a bottom-up approach. For instance, the authors in
\cite{al-shaer06} propose a set of algorithms to detect policy
anomalies in both single- and multi-firewall configuration setups. In
addition to the discovery process, their approach also attempts an
optimal insertion of arbitrary rules into an existing configuration,
through a tree based representation of the filtering criteria.
Nonetheless, we consider their approach as incomplete. Their discovery
approach is not complete since, given a single- or multiple-component
security policy, their detection algorithms are based on the analysis
of relationships between rules two by two. This way, errors due to the
union of rules are not explicitly considered (as our approach
presented in \cite{esorics,safecomp} does).

Although in \cite{al-shaer05} the authors pointed out to this
problematic, claiming that they break down the initial set of rules
into an equivalent set of rules free of overlaps between rules, no
specific algorithms have been provided for solving it. From our point
of view, the proposal presented in \cite{fireman} best addresses such
a problem, although it also presents some limitations. For instance,
we can easily find situations where the proposal presented in
\cite{fireman} reports partial redundancies instead of a single full
redundancy. Moreover, neither \cite{al-shaer06} nor \cite{fireman}
address, as we do in this paper by extending the approach presented in
\cite{esorics,fast}, a folding process for combining both analysis and
refinement strategies.

\section{Conclusions}
\label{sec:conclusions}

\noindent The existence of errors or anomalies in the configuration of
network security components, such as filtering routers or firewalls,
is very likely to degrade the security policy of a system \cite{geer}.
This is a serious problem which must be solved since, if not handled
correctly, it can lead to unauthorized parties to get the control of
such a system.

We introduced in Section~\ref{sec:introduction} two main strategies to
set firewall configurations free of errors. The first approach is to
apply a formal security model --- such as the formal model we presented
in \cite{fast} --- to express the security policy of the access control
for the network, and to generate the specific syntax for each given
firewall from this formal policy --- for instance, by using XSLT
transformations from the formal policy to generate specific Netfilter
configuration rules \cite{netfilter}. A second approach is to apply an
analysis process of existing configurations, in order to detect
configuration errors and to properly eliminate them. In
\cite{esorics,safecomp}, for instance, we presented an audit process based on
this second strategy to set a distributed security scenario free of
misconfiguration.

We presented in Section~\ref{sec:proposal} how to combine both
approaches in order to better guarantee the requirements specified for
a given network access control policy. Thus, from an initial bottom-up
approach, we can analyze existing configurations already deployed into
a given system, in order to detect and correct potential anomalies or
configuration errors. Once verified those setups, we offer to the
administrator a folding mechanism to aggregate the different
configurations into a global security policy to, finally, express by
using a sole formal model, the security policy as a whole. The
security officer can then perform maintenance tasks over such a single
point, and then, unfold the changes into the existing security
components of the system.

As work in progress, we are actually evaluating the implementation of
the strategy presented in this paper by combining both the refinement
process presented in \cite{fast} and the audit mechanism presented in
\cite{esorics,safecomp} (both of them implemented through a scripting
language as a web service \cite{php}). Although this first research
prototype demonstrates the effectiveness of our approach, more
evaluations should be done to study the real impact of our proposal
for the maintenance and deployment of complex production scenarios. We
plan to address these evaluations and discuss the results in a
forthcoming paper.

On the other hand, and as future work, we are currently studying how
to extend our approach in the case where the security architecture
includes not only firewalls but also IDS/IPS, and IPSec devices.
Though there is a real similarity between the parameters of those
devices' rules (as we partially show in \cite{esorics,safecomp} for
the analysis of anomalies), more investigation has to be done in order
to extend the approach presented in this paper. In parallel to this
work, we are also considering to extend our approach to the managing
of stateful policies.\\

\section*{Acknowledgements}

\noindent This work was supported by funding from the French ministry
of research, under the \textit{ACI DESIRS} project; the Spanish
Government (CICYT) projects \textit{TIC2003-02041} and
\textit{SEG2004-04352-C04-04}; and the Catalan Government (DURSI)
grants \textit{2006FIC00229} and \textit{2006BE00569}.

\end{document}